# Hebbian Plasticity for Improving Perceptual Decisions

Tsung-Ren Huang
Department of Psychology, National Taiwan University
trhuang@ntu.edu.tw

**Abstract**

Shibata *et al.* reported that humans could learn to repeatedly evoke a stimulus-associated functional magnetic resonance imaging (fMRI) activity pattern in visual areas V1/V2 through which visual perceptual learning was achieved without stimulus presentation. Contrary to their attribution of visual improvements to neuroplasticity in adult V1/V2, our Hebbian learning interpretation of these data explains the attainment of better perceptual decisions without plastic V1/V2.

Through on-line feedback, Shibata *et al.* (*1*) demonstrated the ability of humans to induce a stimulus-associated functional magnetic resonance imaging (fMRI) voxel pattern in the early visual areas V1/V2. Furthermore, visual sensitivity to an oriented grating stimulus in a noisy background was improved through repeated induction of such a voxel pattern. Interestingly, no grating-related perceptual experience was evoked in their participants by this perceptual learning paradigm. How can stimulus-specific perceptual improvement be attained without external presentation and internal imagery of the stimulus? What exactly has been learned by the neurofeedback participants?

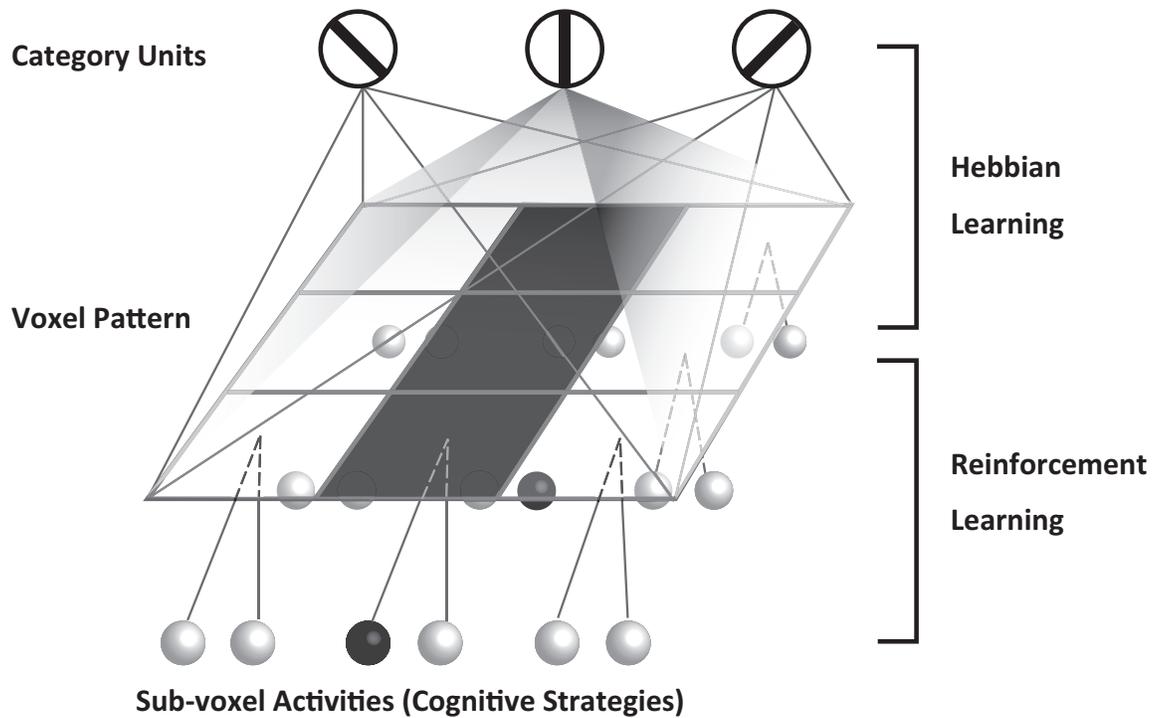

**Fig. 1.** Two proposed types of learning involved in the neurofeedback paradigm. In this schematic, all voxel cells are connected with all category units, and their relative strength of connection weights are learned through Hebbian mechanisms. For reinforcement learning, different cognitive strategies and sub-voxel activities may activate the same pattern at the voxel level.

Our explanation for these data is that both reinforcement and unsupervised Hebbian learning are involved in their neurofeedback procedure (Fig. 1). The reinforcement signal was a variable-size disk indicating the degree of match between a spontaneous voxel pattern in a trial and a stimulus-associated template pattern. The induced template-like patterns in V1/V2 then allowed unsupervised Hebbian mechanisms to strengthen the synaptic connections between repetitively activated V1/V2 neurons and their receiving neurons, such as those in V3. In effect, exposure-specific perceptual decisions were improved because neurons in downstream areas learned to respond more to regular grating signals and to ignore irregular background noises in the inputs. Owing to the limited spatial resolution of fMRI, the possibility remains that different cognitive strategies and neurons at the sub-voxel level could activate the same voxel pattern. Therefore, induction of a grating-associated voxel pattern did not necessarily trigger perception of that grating.

This interpretation, contrary to the conclusion of Shibata *et al*., attributes visual improvements in the neurofeedback study to neuroplasticity in brain structures other than V1/V2. Note that their stimulus-associated template pattern was a voxel pattern obtained *before* perceptual learning, and their reinforcement procedure in fact encouraged preservation of such a pre-training stimulus representation in V1/V2. Because unchanged neural representations cannot account for changed behavioral performances, the observed improvement in perceptual decisions likely reflects neuroplasticity somewhere else in the visual system, such as areas V3 and V4, whose response properties were free to change during the V1/V2 pattern induction stage.

In general, visual perceptual learning (VPL) can result from changes of neuronal responses in early visual areas V1/V2 (*2-4*) and/or higher brain areas that underlie visual recognition, attention and decision making (*5,6*). Although the neurofeedback results support for the latter case from the viewpoint of Hebbian learning, our suggested Hebbian plasticity between connecting neurons may, in principle, occur within or across any brain areas, including V1/V2, in other VPL tasks. Moreover, changes of neural connectivity and response in a neuronal network are often two faces of the same coin—neither characterizes VPL. For example, changed connectivity between the lateral geniculate nucleus and V1 alters V1 neuronal response in an orientation discrimination task (*4*), and changed connectivity between the middle temporal and lateral intraparietal (LIP) areas alters LIP neurononal response in a motion discrimination task (*6*).

To illustrate Hebbian plasticity for VPL, pattern-specific improvement in perceptual decisions was simulated by a two-layer neural network with divisive/shunting normalization in each layer (*7*). The network matches an input pattern $\bar{x}$ with each category prototype $\bar{w}_j$ and outputs the degree of match $y_j$ to category units:

$$y_j = \bar{w}_j \cdot \bar{x} .$$

The best match category *J* is chosen as the perceptual decision, and its category prototype then updates to accommodate the current instance based on an unsupervised Hebbian learning algorithm known as the Instar rule (*8*), or conditional principal component analysis (*9*):

$$\Delta \bar{w}_J \sim y_J (\bar{x} - \bar{w}_J) .$$

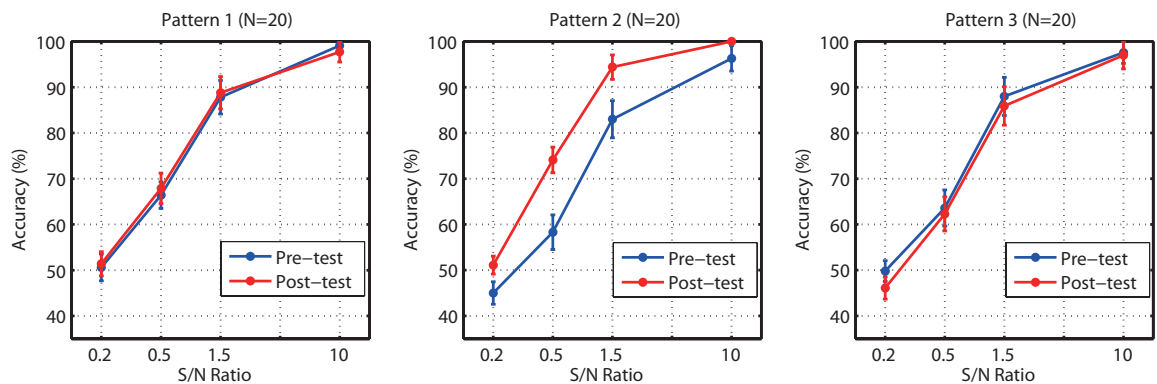

**Fig. 2.** Simulation results of improved perceptual decisions after unsupervised Hebbian learning [cf. Fig. 3A in (*1*)]. Blue and red points indicate classification performances of a 3x3 input pattern with different signal-to-noise (S/N) ratios before and after Hebbian learning. Classification accuracy only improves for a repeatedly presented signal pattern, as shown in the middle panel. Error bars depict the standard errors of the group means from 20 independent simulations with randomized initial weights for network connectivity and randomized presentation order of stimuli to the network.

After learning, a category prototype is approximately equal to the weighted average of category instances, with weights being the degree of match. In other words, the network incrementally learns the regularities in the input patterns, particularly from the low-noise instances. The simulation results (Fig. 2) for a three-alternative forced-choice task qualitatively resemble the behavioral performance reported by Shibata *et al.* in that decision accuracy is a monotonic function of signal-to-noise ratio, and it improves only for the choice corresponding to a repeatedly presented signal pattern. Note that such a Hebbian model can also account for VPL with feedback by adding supervised decision signals into the category units. Thus, it is a general model for learning in perceptual decision making with multiple choices and goes beyond VPL models for binary decisions [e.g., (*4*, *10-11*)].

For the neurofeedback study, whereas the "perceptual" learning component is explicable, the reinforcement-learning (RL) problem faced by participants is challenging in theory due to its infinite state and action spaces. Unlike conventional RL problems with delayed rewards, the neurofeedback study provided immediate evaluative feedback in each trial. The challenge for the participants was to discover effective actions by trial and error before reward-bearing actions could be further exploited. On balance, sufficient exploration is the key to success in reinforcement learning (*12*).

Because participants had no access to the state/voxel space, to optimize reward they had to estimate reward as a function of an experimentally unrestricted, infinite set of action options. In reality, the participants might randomly or parametrically explore a finite set of $N$ actions. Because the overall reward function and its derivatives were unknown with respect to the $N$ actions under exploration, this optimization process required sampling of the $N$-dimensional continuous-valued action space and was expected to be slow with a large $N$ (*13*). Given that action evaluation was sequential in the neurofeedback study and $N$ might dynamically increase when the chosen actions were reward-irrelevant, it is unclear how participants could quickly learn to induce V1/V2 activity patterns within 180 attempts (i.e., the total trial number of a pattern induction session).

In summary, we offer a theoretical perspective on the neurofeedback study by Shibata *et al.*, who substituted stimulus-derived biofeedback signals for real stimuli in visual perceptual training. The lack of stimulus perception might result from other stimulus-unrelated sub-voxel activities that shared the same template pattern at the voxel level. Together with the procedure of clamping a pre-training pattern, it also suggests the observed improvement of perceptual decisions to be a refinement of inference about the input signals rather than the stimulus representation in V1/V2 per se. Such an inference improvement may be attributed to synaptic plasticity in Hebbian learning, which is a viable mechanism for other forms of perceptual learning (*10, 14*). Finally, it remains an open question why rapid learning of pattern induction was observed in the neurofeedback study given that uninformed, sequential exploration of an infinite state/action space is theoretically a slow process.